\documentclass{aa}
\usepackage{graphics}
\begin{document}
\thesaurus{02.18.5;                      
            03.09.04 ;
	    08.14.1;               
            08.16.7                 
              }
\title{High Speed Phase-Resolved 2-d $UBV$ Photometry of the Crab pulsar
 \thanks{Based on observations taken at SAO, Karachai-Cherkessia, Russia}}
\author{A. Golden \inst{1}, A. Shearer \inst{1}, R. M. Redfern \inst{1}, G. M. Beskin \inst{2}, 
S. I. Neizvestny \inst{2},  V. V. Neustroev \inst{2}, V. L. Plokhotnichenko \inst{2} and 
M. Cullum \inst{3}}
\offprints{A. Golden, agolden@itc.nuigalway.ie}
\institute{National University of Ireland, Galway, Ireland 
\and 
Special Astrophysical Observatory, Nizhnij Arhyz, Karachai-Cherkessia, Russia 
\and European Southern Observatory, Garching-bei-M\"{u}nchen, Germany}

\date{Received 10/5/2000; Accepted 3/10/2000}

\maketitle
\markboth{Golden et al., High Speed 2-d $UBV$ Photometry of the Crab pulsar}{}
\begin{abstract}

We report a phase-resolved photometric and morphological analysis 
of $UBV$ data of the Crab pulsar obtained with the 2-d TRIFFID high 
speed optical photometer mounted on the Russian 6m telescope. 
By being able to accurately isolate the pulsar from the nebular 
background at an unprecedented temporal resolution 
(1  $\mu$s), the various light curve components were 
accurately fluxed via phase-resolved photometry. Within the 
$UBV$ range, our datasets are consistent with the existing 
trends reported elsewhere in the literature. In terms of 
flux and phase duration, both the peak Full Width Half 
Maxima and Half Width Half Maxima decrease as a 
function of photon energy. This is similarly the case 
for the flux associated with 
the bridge of emission. Power-law fits to the various 
light curve components are as follows; $\alpha$ 
= 0.07 $\pm$ 0.19 (peak 1) , $\alpha$ = -0.06 $\pm$ 0.19 (peak 2) 
and $\alpha$ = -0.44 $\pm$ 0.19 (bridge) - the uncertainty
here being dominated by the integrated CCD photometry used
to independently reference the TRIFFID data. Temporally, the 
main peaks are coincident to $\le$ 10 $\mu$s 
although an accurate phase lag with respect to the 
radio main peak is compromised by radio timing uncertainties. 
The plateau on the Crab's main peak was definitively determined to 
be $\leq$ 55 $\mu$s in extent and may decrease as a
function of photon energy. There is no evidence for 
non-stochastic activity over the light curves or 
within various phase regions, nor is there evidence of 
anything akin to the giant pulses noted in the radio. Finally, 
there is no evidence to support the existence of a reported 60 
second modulation suggested to be as a consequence of free 
precession.
\keywords{Pulsars: individual (PSR B0531+12), Stars: neutron, Radiation Mechanisms:
non-thermal, Instrumentation: photometers}

\end{abstract}

\section{Introduction}

As the youngest known isolated rotation powered neutron star,
and having the highest known spin-down flux density, the Crab pulsar is 
not surprisingly the most efficient source of high energy magnetospheric emission.
All theoretical models attempting to elucidate the processes of
nonthermal emission must satisfy the empirical constraints provided 
by the extensive observational datasets available for this object
from $\gamma$-rays to the infrared.

Contemporary theoretical frameworks in existence to explain the nonthermal
emission from this and other isolated rotation-powered neutron stars may
be broadly divided into two schools. The first places the source of
emission close to the polar cap region, e.g. Sturrock (1971), and the second 
that place the source at a considerable distance above the neutron 
star surface, in the outer magnetosphere (e.g. Cheng, Ho \& Ruderman 1986). 
In recent times, the development of these models has been to 
some extent restricted to the problem of $\gamma$-ray emission.
These efforts have been predominantly analytical in nature, as in this 
energy regime primary emission is being sampled, which 
avoids complications associated with the lower energy sources of emission.
In these latter cases, the emission is secondary in nature and
a result of interactions that require numerical representation.
Thus the polar cap school (e.g. Daugherty \& Harding 1982) concluded with 
correct flux estimates but poor light curves, and the outer gap
school (Cheng, Ho \& Ruderman 1986) with reasonable light curves and 
approximate flux distributions but requiring extraordinary lines-of-sights 
and magnetospheric physics to justify the emission.
 
The group at Stanford University lead by R. Romani is generally credited by 
being the first to rigourously put the existing models to the test numerically. 
Their conclusions in many ways matched the intuitive premonitions regarding the two model
frameworks, with perhaps the `evolved' outer gap model of Cheng, Ho \& Ruderman (1986)
being the more likely candidate (Chiang \& Romani 1992; Romani \& Yadagiroglu 1995). 
Recently Romani (1998) has indicated that it is only by both the development of 
more advanced numerical simulations that incorporate the full emission physics 
(from $\gamma$-rays to the IR) in a truly self-consistent manner and the 
accurate characterisation of pulsar light curves in the lower energy 
regimes that one can hope to disentangle the problem of nonthermal 
pulsar emission.

Optically, spectroscopic, integrated and high speed single-pixel 
photometry have produced some critically important results in this regard. 
However, true phase-resolved data acquisition with accurate isolation
of the pulsar's emission from that of the nebula requires high
time resolved ($\sim$ 1 $\mu$s) 2-d photometry to confirm 
unambiguously earlier conclusions and to determine the 
true nature of the magnetosphere's emission in the range $\sim$ 1 - 10 eV.
 
In a previous paper (Golden et al. 2000), we presented an analysis
of such observations of the Crab pulsar in which we isolated and theoretically
speculated upon the unpulsed component of the pulsar's optical emission.
These observations were made in January 1996 with the TRIFFID 2-d high speed 
photometer in three colourbands ($UBV$) using the 6m BTA of the Special 
Astrophysical Observatory in the Russian Caucasus. Here we present a
thorough photometric and temporal analysis of the same data. 
Following a review of the pulsar's relevant empirical and theoretical 
characteristics, we briefly outline the observational and data reduction 
methodology - the reader is encouraged to consult Golden et al. (2000) for
a more rigorous treatment of this process. We then detail the results of 
the photometric and temporal analysis, and conclude with a discussion on 
the implications of these results for current thinking on this pulsar, 
with wider ramifications for pulsar emission 
theory in general.

\section{Review}

The Crab pulsar is identified as the south-west member of the
double stars located in the central region of the Crab Nebula
(Cocke et al. 1969).
Radio observations, initially by Staelin $\&$ Reifenstein (1968), 
and regularly ever since, have provided perhaps one of the most 
detailed pulsar rotational datasets to date. 
The pulsar has a 
$P$ $\sim$ 33 ms and $\dot{P}$ $\sim$ 4.211$ \times10^{-13}$
s$\rm s^{-1}$ suggesting a canonical age of $\tau$ $\sim$ 1260
years and surface magnetic field $B_{surface}$ $\sim$ 3.7
$\times 10^{12}$ Gauss.

Historically, high speed optical photometry has been generally 
dominated by the use of single pixel detectors, with variable time
resolution. 
Following the detection of Cocke et al. (1969), numerous targetted
observations followed (Kristian et al. 1970; Wampler et al. 1969;
Warner et al. 1969; Becklin et al., 1973; Muncaster and Cocke, 1972;
Cocke and Ferguson 1974; Groth 1975a,b) with the pulsar
generally observed in the $BVRI$ wavebands at typical
accurate time resolutions of order $\sim$ ms, and the datasets `added'
together in a least-squares fashion - there being no common stabilised
`clock' from which to time the observations, nor a definitive radio 
ephemeris with which to accurately obtain {\it in modulo} 
lightcurves at that time.

The development of more "temporally consistent" instrumentation
in conjunction with other advances in technology allowed for
more detailed observations to be taken.
Peterson et al. (1978) applied for the time rather novel techniques
in the image processing of data obtained via the use of a 6.2 ms
time resolved 2-dimensional Image Photon Counting System camera. 
Significant emission from
the unpulsed presumed `off' component was noted 
in the $U$ and $B$ bands, with Peterson et al. estimating that 
it was some 2\% of the total pulsed emission.

At an effective
time resolution of 20 $\mu$s, Smith et al. (1978) were able to
make unprecedented observations via a S20 photocathode of the
Crab's main pulse region, indicating that the peak had a plateau of
emission $\leq$ 100 $\mu$s in duration. Photon binning at
1 kHz allowed for phase resolved tests to be made of the counting
statistics incident on the detector. They yielded statistical
distributions that did not deviate significantly from random
fluctuations expected, making an allowance for variations in
sky transparancy.

Subsequent high speed observations incorporating an appropriate
polarisation set-up yielded phase resolved polarimetry
(Jones et al. 1981, Smith et al. 1988). 
Under analysis and corrected for interstellar
polarisation, these datasets show evidence for sharp swings in
polarisation angle around both principle peaks, in addition
to some form of evolution of the polarisation in the bridge area.
Remarkably, the results indicated substantial polarisation associated
with what was always assumed to be `off' or at the very least,
the pulsar at a minimum, as noted earlier by Peterson et al. (1978). 


The High Speed Photometer (HSP) on board the Hubble Space Telescope ($HST$) 
was used to observe the Crab pulsar in October 1991 and again in
January 1992, as reported by Percival et al. (1993).
The passbands used were F400LP and F160LP, which crudely approximate 
to standard $V$ and $UV$. 
Although the on-board clock 
is accurate to 10.74  $\mu$s, accuracy to UTC was typically 
$\sim$ $\pm$ 1.05 ms per observing run. Each resulting light 
curve had an effective temporal resolution of 21.5 $\mu$s. 

The pertinent results of Percival et al. (1993) are as follows.
They estimated that the main pulse in the visible leads the radio by 
-1.47 $\pm$ 1.05 ms, and for the UV, -0.91 $\pm$ 1.05 ms, 
with both $UV$ and $V$ main peaks offset by 0.56 
$\pm$ 0.92 ms. Via autocorrelation and least-squares analysis
of the light curves, the Full Width Half Maximum (FWHM), 
peak separation and peak ratios were determined. 
Despite the summing and normalising of the individual light 
curves which might be expected to introduce an element 
of bias in any analysis, Percival et al. (1993) obtain
consistent and reasonable estimates for these parameters.
The implication is that there is a trend whereby the pulse
peaks `tighten' and come together as a function of energy.
Resolution of the main peak cusp was attempted via a
least-squares polynomial fit, and was found to be $\leq$ 150 $\mu$s in
extent. Despite a number of caveats outlined previously 
Percival et al. (1993) can claim to have shown to a reasonable degree that

\begin{enumerate}
\item The peaks' FWHM and separation contract as a function of energy.
\item The $UV$ and $V$ main peaks appear to be coincident to within $\sim$ 1 ms.
\item The $UV$ and $V$ main peaks show signs of leading the radio main peak.
\item A plateau of emission is not definitively resolvable to within 150 $\mu$s.
\item The photon statistics would appear to be Poissonian throughout the light curve.
\end{enumerate}

Later observations, using the $HSP$ in Polarisation mode,
were made of the pulsar (Smith et al. 1996).
Qualitatively the results were similar to earlier optical observations.
The conlusions reached were no different to the original 1988 paper,
other than that any model of the emission regions must be essentially
wavelength independent.

Following on from $IR$ band observations of Ransom et al. (1994),
Eikenberry et al. (1996) made observations in the $J$, $H$ and $K$ bands
using a single-pixel Solid State Photomultiplier, set-up to record 
using 20 $\mu$s time bins. Light curves were then obtained in the
usual manner.

The analysis proceeded along three avenues. The first was determination of
the Half Width Half Maximum (HWHM), essentially a truncated FWHM,
with each side of the FWHM bisected by the maximum of the peak in
question, yielding a `leading' and `trailing' HWHM component.
The second was the determination of phase resolved spectra, and the
third an in-depth examination of the colour variation around the main peak.
Monte Carlo simulations were used to build up a statistical picture of
the optimum fits to the light curve data.

This analysis showed clear evidence that the peaks (morphologically
and in terms of flux) change as a function of energy throughout the
light curve, and also that time limit over which such rapid spectral
changes occured was $\leq$ 180 $\mu$s. Assuming emission occurs in
the vicinity of the light cylinder, then one can estimate the typical
coherence dimensions for the region, being less than or equal to
1.8 $\times 10^{-4}$ s $\times$ $c$ = 54km - indicating that the emission
region is consistent with a localised region of the magnetosphere. 

In a subsequent paper (Eikenberry \& Fazio 1997) these techniques were applied 
to a range of light curves, from $\gamma$-rays (Ulmer et al. 1995), X-rays 
(reconstructed from the $ROSAT$ archives), $UV/V$ (Percival et al. 1993) 
and $KHJ$ (from previous work). The overall energy range spanned 0.47 
eV to 10 MeV. The principle intention was to further test the 
hypothesis that the actual peaks themselves altered morphologically 
with energy. The actual mechanics of how the datasets were analyzed 
was similar to the earlier analysis of Eikenberry et al. (1996) 
- it will suffice to note their more pertinent conclusions;

\begin{enumerate}
\item The ratios of the flux of the bridge and peak 2 to peak 1 show an overall trend to increase
with energy, but reverses at low energies ( $<$ 1eV ).

\item The peak-to-peak separation shows a `nonlinear' trend to decrease with energy over the range,
which is consistent with energy stratification of the emission region.

\item The peaks FWHM show neither a definite trend nor pattern with energy, yet display
significant variability over the range.

\item There is a strong energy dependency as regards the evolution of the HWHM for leading and
trailing edges of both peaks, with some evidence of a continous functionality over the range - except
at the $IR$ region. The HWHM behaviour explains the rather puzzling FWHM behaviour.

\item Peak 2 appears to change profoundly as regards morphology between the $IR$-optical and
X-ray-$\gamma$-ray range, showing a fast rise and slow fall for the former, and the converse
for the latter. This is not explicable by theory.

\item Many of the pulse shape parameters show maxima or minima at the energies of 0.5-1eV in the
$IR$ regime, suggesting some interesting phenomena occuring in this waveband.
\end{enumerate}

Eikenberry \& Fazio (1997) concluded that the many unusual energy dependent 
morphological phenomena that are observed under this analysis were 
incompatible with current model frameworks - in particular the 
single-pole emitting outer-gap (Romani \& Yadigaroglu 1995). In this model,
the emission originates from a topologically extended region that effectively
maps out a hemisphere about the open field region. Consequently, with the
emission profile a result of viewing geometry and crossing caustics to the
line of sight, one would expect either no evident effect, or
a common effect that should manifest itself in a smooth and
continuous manner at all such energies, as the effect would be essentially
the averaging of many seperate spatially emission sources. These 
expectations contrast with the empirically determined results.

\section{Precession Issues}

Free precession of an isolated pulsar, most likely due to some moment of inertia 
inequality, possibly as a result of irregular distributions of neutron star 
matter on its surface, or perhaps as a result of internal interactions and 
effects, would manifest itself as an appropriate modulation in the observed
temporal signal. Previous reports by groups examining the variations in the
ratios of peak 1 to peak 2 in $\gamma$-rays have generally indicated 
periods of the order of years (Kanbach et al. 1990; Nolan et al. 1993;
Ulmer et al. (1994). Previous reports of something akin to a precessional 
like signal from the Crab pulsar in the radio have tended to suggest a 
precessional period of order months/years (e.g. Jones 1988). 

At higher frequencies, evidence of precession could provide both
direct constraints to the condensed matter equation of state, and
imply the emission of gravitational radiation. It is well known that a 
precessing gravitationally compact body rotating at high speeds would be 
expected to radiate gravity waves, at a frequency twice its rotational 
frequency (e.g. Misner, Thorne \& Wheeler, 1971). Various other emission modes
are theoretically possible. Planned ground based gravitational detectors 
are being designed that could have the capability to detect displacement 
amplitudes of order $h$ $\sim$ $10^{-23}$. However do this, these detectors 
must be `tuned' to a particular tight frequency band for up to 1 year to 
build up sufficient $S/N$. Clearly, one needs a guaranteed signal to lock 
on to - and pulsars may very well provide such a source. 

Hence the great interest in the reported suggestions of a precession 
signal with a period of order 60 seconds from the Crab by Cadez and 
Galicic (1996). This was based on a time-series analysis of the High Speed 
Photometer data of Percival et al. (1993), and their own ground based 
observational CCD dataset. Despite nearly three decades of radio 
observations, no such modulation has been reported in the literature.
Furthermore, an analysis of high speed optical data by Jones et al. (1980) 
with early photometers indicated a lack of optical source variations
at the 1 \% level over a ten year period. It is clearly important to 
try and test this with independent data, as a confirmation would have 
important possible implications for other fields, particularly that of 
gravitational wave detection.

\section{January 1996 BTA Run - Observations in $UBV$ $\&$ Data Preparation}

The details of these TRIFFID high speed observations of the Crab pulsar,
and the subsequent preliminary analysis have been published elsewhere 
(Golden et al. 2000). We shall thus limit our description accordingly.
The observations took place over 5 nights in January 1996 using the 
Special Astrophysical Observatory's 6m telescope in the Caucasus. 
TRIFFID high speed camera used incorporated a 2-dimensional photon
counting MAMA detector (\cite{cull90}; \cite{tim85}) with standard
$U$, $B$ and $V$ filters. Timing was accurate to within 400 ns per
10 second period using a GPS receiver. There were 6 $U$ band exposures
totalling 5546 seconds, 5 $B$ band exposures totalling 5750 seconds, 
and 9 $V$ band exposures totalling 6759 seconds. 

\begin{figure}
\resizebox{\hsize}{!}{\includegraphics{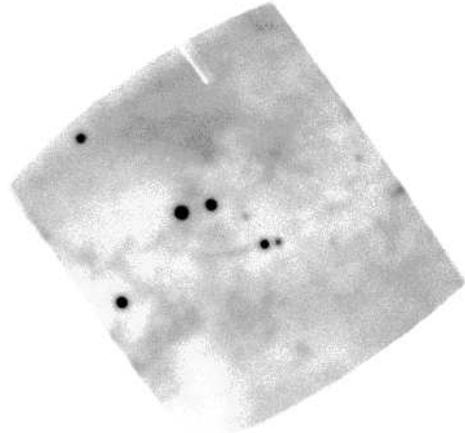}}
\caption{Corrected integrated image for the 96zs2 dataset, in the $V$ band. 
The consequent effect of a dead anode line is visible in the upper part of 
the image. The Crab is the fainter of the two centre-left stars.}\label{96zs2}
\end{figure}

The photon [x,y,t] data stream per dataset was corrected via a Weiner 
filter modified shift-and-add algorithm for telescope wobble and 
gear drift (Redfern et al. 1993). Processing the datasets in this
way and incorporating the derived flat fields yielded full field 
images of the inner Crab nebula from which the pulsar and its 
stellar companions were registered. Figure \ref{96zs2} shows the 
integrated 96zs2.0.0 dataset following this technique. The average 
seeing for this dataset was $\sim$ 1.3$\arcsec$. The Crab pulsar is evident 
as the slightly dimmer upper component of the central double star.

As detailed in Golden et al. (2000), phase resolved images were
obtained by firstly extracting all photons with a fixed pixel
radius of the Crab pulsar, barycentering them, and then 
geometrically aligning each dataset in [x,y] per colour band. 
This yielded a sequence of time-series per colour band that, 
when read by a modified epoch-folding algorithm, generated 
both standard light curves and phase-resolved 2-dimensional 
image within a certain specified phase range. The latter 
used the appropriate data taken from the Jodrell Bank Crab Ephemeris 
(Lyne \& Pritchard, 1996).

\section {Analysis - Temporal Issues}

It is absolutely crucial in this analysis to ascertain the
temporal quality of our datasets, and in this way, test for possible
glitches in the calibrated timing hardware and potential nonlinearities 
caused as a result of microchannel plate saturation, thus 
assessing any unusual temporal activity in the pulsar time series. 

Regarding the problem of $MAMA$ saturation it was determined that,
within the bounds of variable sky transparancy, no non-Poisson behaviour
at the 99\% level was noted, for a range of datasets over the $UBV$
wavebands (Golden 1999). This was repeated throughout the light curve,
indicating that the emission process is consistent with dominant incoherent
synchrotron processes. Giant radio pulses, which are of order 10$^{3}$
times more intense than the average radio pulse, occur in approximately 
1 \% of radio pulse periods (Lundgren et al. 1994). There were no 
deviations in the photon statistics accumulated in these observations
consistent with such giant pulse activity.  

Regarding the temporal stability of the timing system, we previously
determined the timing accuracy to within $\sim$ 1 $\mu$s of 
UTC, and for long (4-5 days) timeseries datasets, it is possible to test the
timing against the Crab pulsar itself via epoch folding or Fourier 
methods. For the former case, this will be as accurate as the timing 
ephemeris provided - during our observing period the Jodrell 
Ephemeris was accurate to 50 $\mu$s. In the latter, we are
restricted to the accuracy with which one can determine to have detected
a harmonic at high $\nu$. Via epoch folding, the best reduced $\chi^{2}$ was
determined for the given rotational parameters from Jodrell within
100 $\mu$s of the given reference time stamp given by the GPS system.
Using an Origin2000 supercomputer, detection of
the high harmonics (typically $\sim$ 50) for several daily $B$ datasets
yielded frequency estimates that were accurate to the corrected ephemeris
frequency to within 1 in 10$^{7}$. Consequently, we were
satisfied with the temporal integrity of these MAMA datasets taken
in Russia in January 1996.

In this vein, the unusual low frequency signal of $\sim$ 0.016 Hz
reported by Cadez \& Galicic (1996) was put to the test. FFT analysis
on each night's dataset binned at 1, 0.1 \& 0.2 Hz were tested, but
no significant signal was apparent within the 0.01 - 1 Hz window -
or at greater frequencies. Consequently we must remain somewhat
guarded as regards the conclusions of Cadez \& Galicic (1996).
The definitive discovery of a
60 second precessional signal from an isolated rotating neutron star
would have profound implications for fundamental physics and thus
demands conclusive and undeniable evidence for its justification.
We must state that we see no such evidence to date.

\section {Analyses - Phase Resolved Photometry}

\begin{figure}
\resizebox{\hsize}{!}{\includegraphics{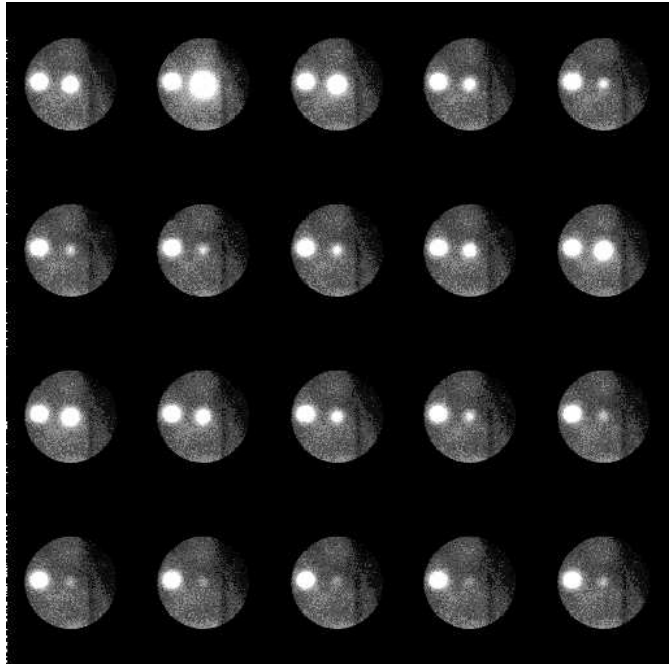}}
\caption{Mosaic image of one cycle in $V$ of the Crab Pulsar. The sequence
of phase-resolved frames starts at the top left. To `read' the frames,
one moves from left to right along each row. This continues until
one reaches the bottom right frame. Thus one can
see the main peak initially, the bridge of emission (row 2), the
interpulse/peak 2 (end of row 2, row 3), and the `off' phase (row 4).
The structure on the right hand side is the dead anode line as seen in 
Figure 1, and its location in proximity to the pulsar occurred 
in only one dataset.}\label{mos_crab}
\end{figure}

By the use of developed epoch folding algorithms, it was possible to take the 
deep summed time series for all three wavebands and re-construct a 
phase-resolved image for a specific phase region. Figure \ref{mos_crab} 
shows such an {\it ad hoc} mosaic of phase-resolved images, with the phase 
resolution here 0.05 of phase.
It is apparent that there appears to be residual emission from the Crab pulsar
during the traditional `off' phase (essentially row 4 in Figure 2), as was 
originally noted by Peterson et al. (1978). The fact that we can obtain deep 
phase-resolved images which have been corrected for telescope drift and 
wobble with a sufficient field of view enables us to rigorously characterise 
the time-resolved nature of the pulsar's emission via standard image 
processing techniques.

There are a number of ways to proceed with this analysis. Principally, we wish to determine
1) what the total background is in a given aperture per colour band and 2) what are the
respective fluxes for phase-resolved  components of emission associated with the Crab pulsar.
The first step is perhaps the most important as it may then be applied  to the standard
derived light curves so as to correct for the background per bin and thus allow one to
characterise the so-called `steady' or `off' phase of emission.

In Golden et al. (2000), the image processing techniques were outlined in some
considerable detail for this analysis. It is sufficient to state that the
IRAF {\tt daophot} package was used to remove the full Crab image (and thus
obtain the background per aperture per colour band in addition to the
full Crab flux), and similarly for the various phase-resolved light curve 
components, which are localised in terms of phase in Table \ref{eik_comps}.
This process involved the empirical fitting of a PSF 
to the Crab and making the subsequent extraction. 
\begin{table}
\caption[]{Pulse Phase Definitions.}\label{eik_comps}
\begin{center}
\begin{tabular}{cl}
\hline\noalign{\smallskip}
  Phase     	&	Region  \\
\hline\noalign{\smallskip}
Peak 1  		&  -0.12 - 0.12   \\
Peak 1 (Leading Edge) 	&   -0.12 - 0.0  \\
Peak 1 (Trailing Edge)  &    0.01 - 0.12 \\
Bridge 			&    0.121 - 0.27 \\
Peak 2 			&    0.271 - 0.658 \\
Peak2 (Leading Edge) 	&    0.271 - 0.409 \\
Peak2 (Trailing Edge) 	&    0.410 - 0.658 \\
Off Phase  		&  0.75 - 0.825   \\
\hline\noalign{\smallskip}
\end{tabular}
\end{center}
\end{table}

The `off' phase here was localised in the following manner. Correcting
for the estimated background above, the `off' regions for the various 
coloured light curves at 11 $\mu$s resolution (3000 bins) were 
tested via a $\chi^{2}$ algorithm. Here, the bin region was iteratively
reduced, and tested to see if the observed flux was consistent with
that expected from a non-varying source over that phase region, at
the 95 \% level.

As regards any nebular contribution to this photometric analysis, no 
significant change was noted between the estimated flux of the `off phase' 
made via extraction using the full cycle PSF, and the flux determined by 
empirically fitting a PSF to the `off phase' itself. 
This indicates that the local nebulosity does not significantly 
contribute to the determined fluxes in this work.

\subsection {Phase Averaged/Integrated Colour Photometry}

Actual phase-resolved photometry was done by normalizing the total
Crab flux per colour band to a known integrated photometric reference,
and then determining the contribution per phase region based on the
fractional fluxes estimated for that particular phase-resolved component.
The extinction corrected $UBV$ ground-based photometry of Percival et al. (1993)
was used as the reference data for this purpose

Table \ref{eik_anal3} displays the tabulated fractional fluxes estimated following
the photometric analysis detailed previously within each colour band - the fluxes
were derived via renormalizing the fractional flux with respect to the
Percival et al.'s integrated data. 
\begin{table*}
\begin{center}
\caption[]{Fractional Flux derived from Photometric Analysis.}\label{eik_anal3}
\begin{tabular}[t]{cccc}
\hline\noalign{\smallskip}
  Parameters	&			  & Waveband		   &		    \\
                 	 &   U  (mJy)    		  &  B  (mJy)    		   &  V  (mJy)	    \\
\hline			
\noalign{\smallskip}
\hline\noalign{\smallskip}
Peak 1                  & 2.0   $\pm$ 0.1    & 1.9   $\pm$ 0.1  & 1.9  $\pm$ 0.1 \\
Peak 2                  & 1.1   $\pm$ 0.1    & 1.1   $\pm$ 0.1 & 1.1 $\pm$ 0.1 \\
Bridge                  & 0.09  $\pm$ 0.01   & 0.10 $\pm$ 0.01  & 0.11 $\pm$ 0.01 \\
Off Phase               & 0.014  $\pm$ 0.002 & 0.017 $\pm$ 0.002 & 0.019 $\pm$ 0.002 \\
P1 (HWHM) Lead          & 1.1 $\pm$ 0.1    & 1.2 $\pm$ 0.1    & 1.1  $\pm$ 0.1 \\
P1 (HWHM) Trail         & 0.8 $\pm$ 0.1      & 0.8 $\pm$ 0.1    & 0.8  $\pm$ 0.1 \\
P2 (HWHM) Lead          & 0.46  $\pm$ 0.03   & 0.49 $\pm$ 0.03  & 0.46  $\pm$ 0.03 \\
P2 (HWHM) Trail         & 0.59 $\pm$ 0.04    & 0.65 $\pm$ 0.04  & 0.65  $\pm$ 0.04 \\
\hline\noalign{\smallskip}
\end{tabular}
\end{center}
\end{table*}

\begin{table*}
\begin{center}
\caption[]{Estimated Spectral Power-Laws from Photometric Analysis}\label{slopes}
\begin{tabular}{cc}
\hline\noalign{\smallskip}
Dataset  &   Power-Law            		\\
         &     $\propto$ $\nu^{\alpha}$         \\
\noalign{\smallskip}
\hline\noalign{\smallskip}
Integrated UV/U/B/V/R   &  0.11 $\pm$ 0.09      \\
Integrated U/B/V        & -0.07 $\pm$ 0.19      \\
Peak 1                  &  0.07 $\pm$ 0.19     \\
Peak 2                  & -0.06 $\pm$ 0.19    \\
Bridge                  & -0.44 $\pm$ 0.19      \\
Off                     & -0.60 $\pm$ 0.37     \\
Peak 1 HWHM (Leading)   & -0.15 $\pm$ 0.19    \\
Peak 1 HWHM (Trailing)  & -0.11 $\pm$ 0.19     \\
Peak 2 HWHM (Leading)   & -0.04 $\pm$ 0.19      \\
Peak 2 HWHM (Trailing)  & -0.18 $\pm$ 0.19     \\
\noalign{\smallskip}
\hline
\end{tabular}
\end{center}
\end{table*}

In Table \ref{slopes}
the estimated power-law parameter $\alpha$ has been determined via a weighted 
linear least-squares fit for each individual spectral dataset, with associated 
errors. We have re-calculated $\alpha$ for the both the full range ($UVUBV$) 
\& $UBV$ ground-based Percival et al. (1993) datasets to compare with 
the other power-law fits. Included in this Table are power-law fits to the
FWHM and HWHM components.

\subsection {Phase Resolved Colour Spectra - the Peaks \& Bridge}

\begin{figure}
\resizebox{\hsize}{!}{\includegraphics{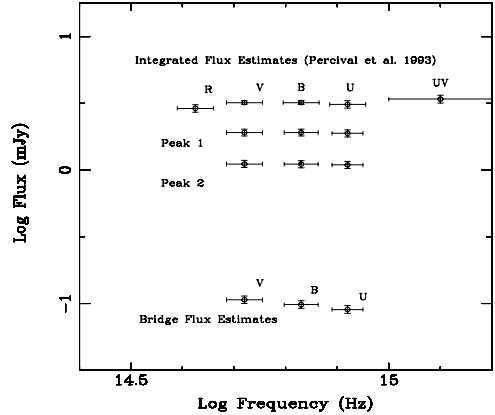}}
\caption{Phase-resolved Fluxes over $UBV$ for peaks 1 \& 2 and the bridge component of emission}\label{peaks_bridge_fluxes}
\end{figure}

\begin{figure}
\resizebox{\hsize}{!}{\includegraphics{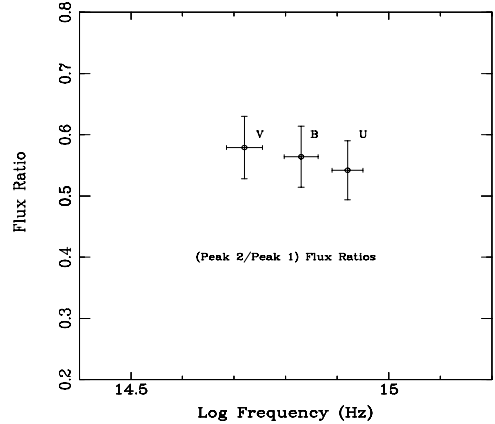}}
\caption{Flux Ratios between peaks 1 \& 2 over $UBV$}\label{peak1_peak2_flxrat}
\end{figure}

\begin{figure}
\resizebox{\hsize}{!}{\includegraphics{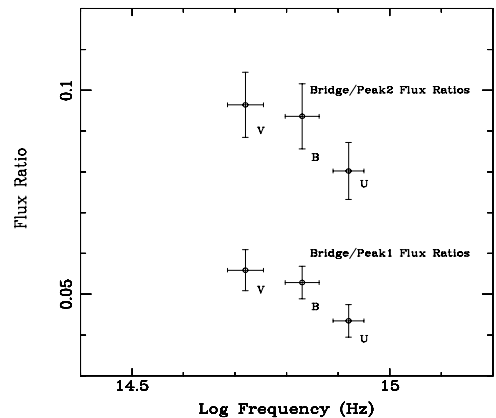}}
\caption{Flux Ratios for peaks 1 \& 2 with respect to the bridge emission.}\label{bridge_peaks_flux}
\end{figure}

Figure \ref{peaks_bridge_fluxes} shows as a function of photon energy the spectral forms
based on the fluxes of each peak, and to the same scale and the emission from the bridge
component.
It seems clear that certainly as regards the bulk flux from both peaks, that
the functional forms, with $\alpha$ $\sim$ 0.07 $\pm$  0.19 for peak 1 and
$\alpha$ $\sim$ -0.06 $\pm$  0.19 for peak 2, are to first order identical with 
that reported for the integrated emission, namely a flat power law
($\alpha$ $\sim$ -0.07  for the ground-based $UBV$, $\alpha$ $\sim$ 0.11 
for the full ground-based $UVUBV$ Percival et al. (1993) dataset). This is not entirely 
surprising, as the dominant emission from the pulsar is localised in 
the peaks, and the fact that both are identical in form suggests a 
common population of electrons/magnetic field environment/Lorentz factor. 
Figure \ref{peak1_peak2_flxrat} substantiates this common source,
there being no significant difference between the colour bands evident 
in these datasets. There appears to be, however, evidence for a somewhat 
steeper power-law associated with the bridge component of emission 
($\alpha$ $\sim$ -0.44 $\pm$  0.19). 
In Figure \ref{bridge_peaks_flux} the fractional bridge/peak fluxes are
shown as a function of photon energy, both suggesting similar functional forms,
consolidating the common proportionality between each flux component.
This suggests to first order a different population of emitting
electrons / magnetic field environment / Lorentz factor  in terms of the 
bridge component than that for the peaks. However it should be stressed 
that any definitive differing power-law
forms are only significant at the 2$\sigma$ level - in this we are at 
the mercy of the reference data of Percival et al. (1993), where the 
substantial error component originates. Using other reference data 
of a higher photometric quality and further observations at other
frequency ranges (such as in the $R$) would constrain these fits 
significantly.

\subsection {Phase Resolved Colour Spectra - Leading \& Trailing Peak Edges}

\begin{figure}
\resizebox{\hsize}{!}{\includegraphics{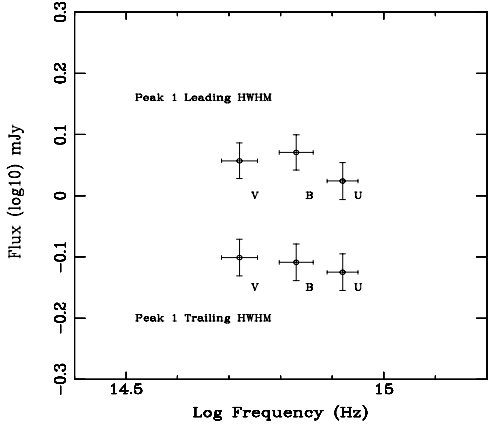}}
\caption{Leading and trailing HWHM Flux Estimates for peak 1}\label{peak1_hwhm_flux}
\end{figure}

\begin{figure}
\resizebox{\hsize}{!}{\includegraphics{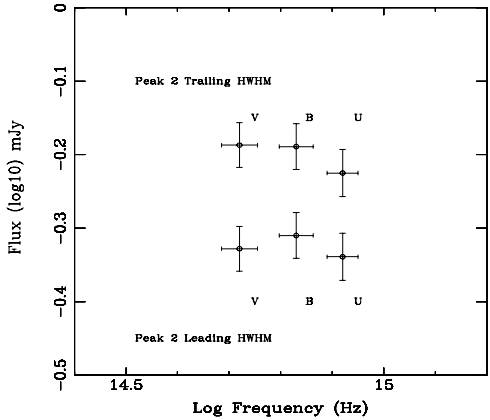}}
\caption{Leading and trailing HWHM Flux Estimates for peak 2}\label{peak2_hwhm_flux}
\end{figure}

In Figures \ref{peak1_hwhm_flux} \& \ref{peak2_hwhm_flux} the HWHM flux
estimates for both peaks 1 and 2 are shown - here the leading and trailing
HWHM fluxes are derived for each peak. It seems clear
that the leading \& trailing edges for both peaks present spectral forms
that are in agreement with each other (see Table \ref{slopes}), although
we are somewhat restricted in our analysis by error considerations.

For both cases, we see suggestions for power-laws with decreasing exponents 
as a function of frequency, although both are consistent with an essentially 
flat functional form. These results are in agreement with estimates of 
Eikenberry \& Fazio (1997), whose analysis yielded results which incline 
towards a decrease in flux with increasing photon energy. It is at $JHK$ 
wavebands that Eikenberry et al. (1996) observed quite profound changes 
in the overall spectral forms for the leading \& trailing edges of 
both peaks - certainly in the $UBV$ regime, we appear to be sampling 
similar synchrotron emitting populations.

\section {Analyses - Morphological Characteristics}

As has been outlined previously, there has not been to date a consistent
set of observational datasets that are temporally accurate (to 1 $\mu$s of UTC),
and this has to some extent compromised previous efforts. The data taken
in January 1996 provides the first such thorough sample, and we consequently
first concentrate on its temporal analysis. 

With an effective estimate of the background obtained for all three colour bands
as previously outlined, one can proceed with a rigourous morphological analysis of the
light curves themselves. This
is essentially pursued by use of both least-squares fits to the Poisson 
`corrupted' light curve profile, and the use of Monte Carlo techniques to 
ascertain the errors on any ideal topological fit to the data. In this way, 
we can hope to obtain information on a given peak's precise maximum position in 
terms of phase, the extent of a plateau within the peak's maximum region, 
and the peak's FWHM and HWHM characteristics.

The basic algorithm fits third order polynomials to the leading and 
trailing edges of the given peak (from 30\% to 90\% maximum intensity), 
and then a sixth order polynomial across the peak cusp (from 90\% to 90\%). 
This latter fit is then used as a template in the following manner. A 
Monte Carlo routine uses the template as a basis from which to
`add' Poisson noise dependent on the template counts per specific bin. 
Having done this, a subsequent sixth order polynomial is fit via 
least-squares. From this, the phase bin with the maximum intensity is 
determined, and `noted'. This maximum phase is then used in the calculation 
of the FWHM and HWHM parameters, using the previously determined
fits to the leading and trailing edges of a given peak.

An iterative routine then
starts with the two adjacent bins to the nominated maximum bin, computes 
the local average per bin, and computes the $\chi^{2}$ to test for a 
deviation from this local average i.e. the end of a `plateau' region. 
The range is extended by a bin on each side, and the process is 
continued until significant deviations are noted at the 95\% and 99\%
confidence intervals.

Finally, within the defined bin regions defined by these
confidence ranges, a final routine starts with 2 bins, and sweeps through the
ranges, noting the $\chi^{2}$ statistic, and incrementing the bin number. The
$smallest$ number of bin sizes that deviate from the required confidence levels are noted.
The entire process is repeated from the initially derived template $10^{4}$ times.
For each chosen parameter, the average is calculated, and the standard error on the
$10^{4}$ samples is used to represent 1$\sigma$ errors on that mean.

The above outlined approach has its origin in the analytical techniques of Ransom,
Eikenberry and others (Ransom et al., (1994), Eikenberry et al. (1996))
with the Monte Carlo derived estimate of a plateau region incorporated.
The completed temporal analysis for the light curves in $UBV$ is tabulated in
Table \ref{eik_anal1}.

\begin{table*}
\begin{center}
\caption[]{Pulse Shape Analysis Results for January 1996 Crab data}\label{eik_anal1}
\label{obs}
\begin{tabular}{cccc}
\hline\noalign{\smallskip}
 Parameter		&			& Waveband		&		    \\
                 	 &   U      		&  B      		&  V 		    \\
                         &   \null  		&  \null  		&  \null            \\
\hline{\smallskip}			
                         &   Phase  		  &  Phase  		   &  Phase         \\

\noalign{\smallskip}
\hline\noalign{\smallskip}
Peak 1 position		&  0.002 $\pm$ 0.003	& 0.003  $\pm$ 0.001     & 0.0024 $\pm$ 0.0003 \\
Peak-to-peak separation &  0.407 $\pm$ 0.003	& 0.405  $\pm$ 0.002     & 0.405  $\pm$ 0.002 \\
Peak 1 FWHM		&  0.043 $\pm$ 0.003	& 0.043  $\pm$ 0.001     & 0.045  $\pm$ 0.000 \\
Peak 2 FWHM		&  0.078 $\pm$ 0.003	& 0.086  $\pm$ 0.001     & 0.089  $\pm$ 0.002  \\
Peak 1 HWHM (lead)	&  0.028 $\pm$ 0.004	& 0.030  $\pm$ 0.001     & 0.030  $\pm$ 0.001 \\
Peak 1 HWHM (trail)	&  0.015 $\pm$ 0.003	& 0.014  $\pm$ 0.001     & 0.015  $\pm$ 0.001 \\
Peak 2 HWHM (lead)	&  0.037 $\pm$ 0.002	& 0.038  $\pm$ 0.001     & 0.039  $\pm$ 0.001  \\
Peak 2 HWHM (trail)	&  0.041 $\pm$ 0.002	& 0.047  $\pm$ 0.002     & 0.050  $\pm$ 0.002  \\
\noalign{\smallskip}
\hline
\end{tabular}
\end{center}
\end{table*}

\subsection {Peak 1 position \& Peak-to-Peak Phase Difference}

\begin{figure} 
\resizebox{\hsize}{!}{\includegraphics{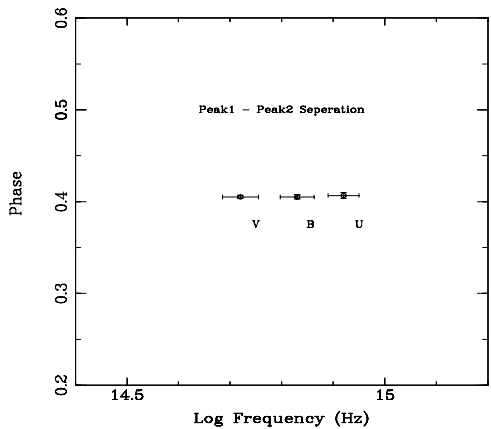}}
\caption{Peak 1 \& peak 2 separation, in units of phase.}\label{p2p_d}
\end{figure}

For each of the two peaks per colourband, as stated above, the iterative 
Monte Carlo routine determines the positions of the `best fitted' 
peak maxima, and collates the difference between the two in phase, 
as well as recording the position in phase of the first peak. 
Figure \ref{p2p_d} displays the peak-to-peak difference as a function 
of frequency. It would seem that there is no clear trend for the peaks 
phase difference to deviate within this restricted photon energy regime.

Based on the
errors associated with our data, we do not see any departure from this trend
reported elsewhere in the literature that the peak-to-peak phase difference
is seen to contract with increasing energy (e.g. Eikenberry et al. 1997;
Ramanamurthy 1994). Similar error concerns restrict any real attempt to
definitively state that the $UBV$ main peaks arrive at differing phase
intervals. The trend, if any, is that the radio-optical main peak difference
in arrival phase increases with photon energy - broadly consistent with
Percival et al. (1993) and other authors.

\subsection {Full Width Half Maximum Differences}

\begin{figure}
\resizebox{\hsize}{!}{\includegraphics{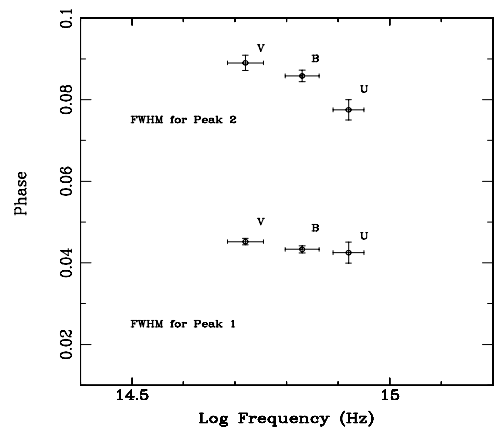}}
\caption{FWHM in units of phase, for peaks 1 \& 2 as a function of $UBV$.}\label{fwhm_ph}
\end{figure}

The FWHM of a pulsar's pulsed components as a 
function of energy have been used for many years as a way of understanding 
the emission characteristics of these objects, being undoubtedly related 
to the geometry of the actual emission mechanism. The expected FWHM 
behaviour with photon energy is thus a strong function of the model
under scrutiny - for example, the standard Outer Gap formalism predicts a
growing FWHM with decreasing photon energy (Eikenberry et al. 1997), and early
rather monolithic models envisaging emission occuring from an open cone
interpret the morphologies as the collected open `cones' of synchtrotron
emitting population of electrons. In this latter case, a similar effect is
expected. Our Monte Carlo results, as tabulated and presented in 
Figure \ref{fwhm_ph}
for both peaks substantiate this trend - although there seems to be
a statistically significant greater gradiant associated with peak 2 compared
with peak 1. From the previous work of Eikenberry et al. (1997) using
the HSP dataset, it is impossible to make such a differentiation.

\subsection {Half Width Half Maximum Differences}

\begin{figure}
\resizebox{\hsize}{!}{\includegraphics{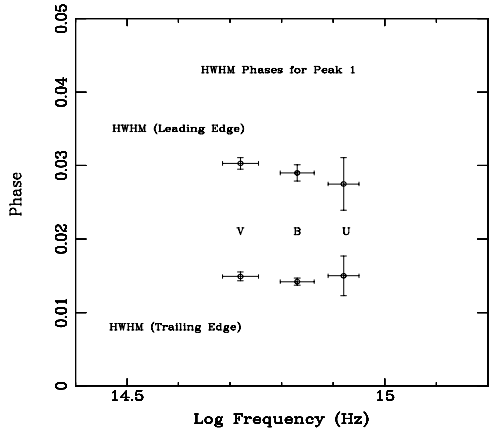}}
\caption{HWHM in units of phase, for the leading \& trailing edges of peak 1.}\label{hwhm1_ph}
\end{figure}

\begin{figure}
\resizebox{\hsize}{!}{\includegraphics{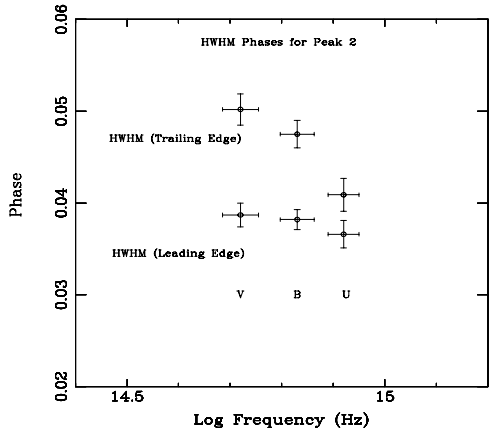}}
\caption{HWHM in units of phase, for the leading \& trailing edges of peak 2.}\label{hwhm2_ph}
\end{figure}

In their analysis, Eikenberry et al. (1997) found statistically significant
differences between the leading and trailing half-widths for both peaks, these
deviations clearly evident at the lowest photon frequencies ($JHK$), with
the trend at our optical wavebands being an increase in measured phase extent
and decreasing photon energy. Figures \ref{hwhm1_ph} and \ref{hwhm2_ph} display the results
following the Monte Carlo analysis (tabulated in Table \ref{eik_anal1}), and there
is agreement with these earlier conclusions within the spread of
errors.
The leading edges for both peaks display similar functional forms.

\subsection {Plateau on the Main Peak}

\begin{figure}
\resizebox{\hsize}{!}{\includegraphics{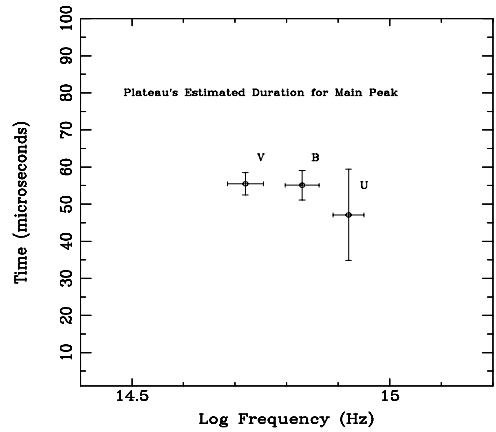}}
\caption{Result of Monte-Carlo Analysis on the Main Peak's temporal extent.}\label{plat_times}
\end{figure}

Resolution of the main peak has been discussed many times in the literature as
a diagnostic as regards the basic emission mechanism for principally the Crab
pulsar, as data with the required S/N is unlikely (at this stage) to come from any
other object; but as outlined earlier, crude temporal resolution and not
sufficiently rigorous analytical techniques have always dogged this. With
the background-subtracted datasets, the same Monte Carlo code was used to
ascertain the extent of non-deviating `best-fit' 6th order polynomial across
the cusp of the main peak with respect to the local noise (at the 95\% and
99\% confidence levels). The results of this approach are tabulated in Table
\ref{eik_anal2}, and in Figure \ref{plat_times} we show the estimated plateau duration
as a function of photon energy.
\begin{table*}
\begin{center}
\caption[]{Peak Plateau Analysis for January 1996 Crab data}\label{eik_anal2}
\label{obs}
\begin{tabular}{cccc}
\hline\noalign{\smallskip}
 Plateau Parameters	&			  & Waveband		   &		    \\
                 	 &   U      		  &  B      		   &  V 	    \\
                         &   \null  		&  \null  		&  \null            \\
\hline{\smallskip}			
                         &   Phase  		  &  Phase  		   &  Phase         \\
\noalign{\smallskip}
\hline\noalign{\smallskip}
Local Average, 99$\%$	&  0.0014    $\pm$ 0.0004 & 0.0017 $\pm$ 0.0001 & 0.0017     $\pm$ 0.0001 \\
Peak Intensity, 99$\%$	&  0.0013    $\pm$ 0.0004 & 0.0016 $\pm$ 0.0001 & 0.0017     $\pm$ 0.0001 \\
\hline\noalign{\smallskip}
                         &    $\mu$s  		  &   $\mu$s  		   & 
			  $\mu$s         \\
\noalign{\smallskip}
Local Average, 99$\%$	&  47.5 $\pm$ 12.7 & 55.1 $\pm$ 4.0 & 55.5 $\pm$ 3.0 \\
Peak Intensity, 99$\%$	&  45.0 $\pm$ 13.7 & 55.1 $\pm$ 0.3 & 55.5 $\pm$ 3.0 \\
\noalign{\smallskip}
\hline
\end{tabular}
\end{center}
\end{table*}
Within the errors, we can say to first order that the cusp of the main peak is
consistent with the presence of a `plateau' of emission lasting some 
$\sim$ 50 $\mu$s
in exent, which is in agreement with the previous conclusions of Komarova et
al. (1996). Tantalisingly, a possible trend of
decreasing plateau exent with increasing photon energy will have to await
further datasets; there do appear to be indications that this may be the case
from our data.

\section {Discussion}

Putting it succinctly, despite the excellent temporal resolution of our
data, our analysis and subsequent conclusions are limited by both
the low throughput of the imaging system and the large errors associated with the
reference integrated photometry of the Crab pulsar.
Nevertheless, we have resolved to an
unprecendented photometric accuracy the flux fractions associated
with the principal regions of the light curve, most especially that
of the unpulsed component. Within errors, the datasets confirm
existing trends seen in previous analyses; the intra-peak separation,
peak FWHM/HWHM, and bridge flux as a function of photon energy.
In this, we see nothing unusual to most model predictions. However
the following results provoke further scrutiny.

\subsection{Resolution of the Unpulsed Component of Emission}

We have outlined in some considerable detail elsewhere the
implications of the observed spectral form for the resolved
unpulsed flux component (Golden et al. 2000). Clearly it
is nonthermal, and spectrally similar to that of the bridge
of emission. In our previous communication we argued
that the observed unpulsed component of emission has its 
source in a similar electron population/magnetic field/Lorentz 
factor environment, but stated that an unequivocal link was
not evident, and that viewing geometry was likely to play a
substantial role in what we ultimately observe.

\subsection{The temporal extent of the Main peak's cusp}

Resolution of the main peak's cusp temporally has long been regarded as
a possibly powerful diagnostic as regards determining the local physical
conditions where the emission is occuring, which in the optical regime
is believed to be as a result of incoherent synchrotron processes.
In this analysis we have obtained rigorous estimates of the
extent of this plateau, namely $\leq$ 55 $\mu$s with some indication
of an increase with wavelength. It now remains to place this
result in some sort of theoretical context. The models
imply emission either occuring from a spatially extended region
(the Outer Gap of Romani \& Yadigaroglu (1995) or from 
a localised cone/`trumpet' like zone traced by the 
opening angle above a polar cap, and subsequently
modified by the toroidal dipolar field. Thus for the former,
the peaks are a consequence of many photons arriving from disparate
locations but in phase due to relativistic effects, and in the latter
we have a conceptually more appealing core/cone of emission.
In both cases, the synchrotron emitting electron's are expected
to have a common Lorentz/local magnetic field, in order to
satisfy the observed luminosity, although the intuitive concept of a
`beam' is problematic for the former $ansatz$.

For a temporal extent of 55 $\mu$s at the main peak, one could argue that
this implies that the dominant beaming from the population of synchrotron
emitting electrons from some localised area in the magnetosphere has a
total angular scale of $\leq$ 10$^{-1.98}$ radians.
It follows (Sturrock et al. 1975) then that if one envisages that the total
optical emission comes from these electrons with a Lorentz factor
$\gamma$ and pitch angle $\xi$, then
\begin{equation}
\gamma \geq \gamma_{o} = 10^{1.98}, \; \; \xi \leq \xi_{o} = 10^{-1.98}
\end{equation}
suggesting a $low$ Lorentz factor emission regime ($\sim$ $10^{2}$).
If the radiation is via normal (large-angle) synchrotron emission,
then the emitted spectrum peaks at the frequency
\begin{equation}
\nu = \frac{3}{2} \nu_{G} \gamma^{2} \xi \;  \rm Hz \; \;(\gamma \xi \geq \frac{4}{3})
\end{equation}
where $\nu_{G}$ is the gyrofrequency given by
\begin{equation}
\nu_{G} = 10^{6.5} B \;  \rm Hz
\end{equation}

In the small angle approximation, the spectrum peaks at

\begin{equation}
\nu = 2\nu_{G}\gamma \; \rm Hz \; \;(\gamma \xi \leq \frac{4}{3})
\end{equation}

It is easily shown that for either case,

\begin{equation}
\nu > 2\nu_{G} \gamma_{o} \; \rm Hz 
\end{equation}

From Eikenberry et al. (1997), the optical spectrum appears to
show a maximum in the vicinity of the H band ($\sim$ 1.8 $\times 10^{14}$ Hz)
or $10^{14.26}$. Applying this with the above suggests that the
local magnetic field must be
\begin{center}
\begin{equation}
B \leq 3.02 \times 10^{5} \;  \rm Gauss
\end{equation}
\end{center}

Assuming a dipolar $\propto$ $r^{-3}$ law, and  the canonical
10km neutron star possessing a derived surface magnetic field
of 3.7 $\times$ $10^{12}$ Gauss, this corresponds to a radial
distance of some 2256 km. In comparsion, the $R_{lc}$ is
estimated to be $\geq$ 1523 km - thus emission must occur
at some 1.5$R_{lc}$.
This emission is not possible for the ideally orthogonally aligned
rotator of Smith (1978) and others, but feasible for those pulsars
whose angle $\alpha$ between dipole \& rotational axes is $<$ 90$^{o}$.
It is clear that the assumption of restricted radial zones of
emission would favour the more localised models over
the spatially extended (where zonally varying local
$\gamma$ \& $B$ would contribute to the overall pulse profile)
and that this radial estimate has many intuitive links with
the emission model of Gil et al. (2000).

As a final comment, we note some evidence for the plateau's
temporal extent to grow with wavelength, errors notwithstanding.
This trend, if real, could be possibly explained most easily with
reference to synchrotron self-absorption. Previous discussions
on this topic assumed that the observed roll-over was
due to this phenomenon, yet time-resolved observations of
the $IR$ wavebands by several authors (Middleditch et al. 1983;
Eikenberry et al. 1996, 1997) do not neccesarily show evidence
for a flattening of the main peak (however, Pennypacker 1981; Penny 1982).
This effect should be manifested in the $IR$ wavebands, and it
is somewhat difficult to justify at our low frequencies for a
tightly localised synchrotron source - but it is possible in a radially or 
latitudinally extended model, where differing regions of the magnetosphere
may experience the effect. Alternatively, we can define the cusp as a simple
function of the local synchrotron emission processes. Here, the lower energy
electrons have a wider opening angle, and thus we observe a wider plateau.
Accurate numerical modelling would test this latter
hypothesis.

\subsection{Phase difference between radio \& $UBV$ Main peak arrival}

From the tabulated data (Table 4), we can say for the higher 
S/N data of $B$ \& $V$ that the main peak pulses arrive to within 10 $\mu$s of each other,
and that, when one considers the peak-to-peak separation, there does
not seem to be any clear evidence for variation as a function of
photon energy. In this we are somewhat restricted by the overall poor
bin S/N for these datasets. Deeper observations would provide temporally
more accurate data, but to first order, it does seem that in these
wavebands, we are seeing an essentially constant general behaviour,
as regards pulse arrivals. We cannot absolutely differentiate the
radio/pulse lag time, due to the error on the former (some 50 $\mu$s),
but we can place a relative lag of order $\sim$ -60 $\mu$s - that
is to say, the $UBV$ emission trails the arrival of the radio main
pulse. This might not be so difficult to understand if the radio
emission had its source at some distance further in towards the
neutron star surface with respect to the higher energy sources, located
in the toroidally warped dipolar field, where both relativistic beaming
and the local magnetic field vector might conspire to produce such
a delayed arrival time (assuming localised emission). But this is
not the case for the Crab, whose radio emission is believed to occur
in proximity to the optical sources close to the light cylinder.
Observations with the FIGARO II $\gamma$-ray instrument (Masnou et al. 1994)
have suggested that the (0.15 - 4.0 MeV) $\gamma$-ray main peak $precedes$ that
of the radio pulse by some 375 $\pm$ 148  $\mu$s, which according to the
authors, points to spatially different regions of emission between
the radio \& $\gamma$-ray source regions of some 100 km in the
magnetosphere. Recent RXTE observations (Rots et al. 1998)
at a superior temporal resolution with respect to UTC suggest
that the (5-200 keV) X-ray main peak leads the radio
pulse by 264 $\pm$ 330  $\mu$s.

For our `best' position, the $V$ band, the main
peak is localised at +79.2 $\pm$ 10  $\mu$s with respect to the radio
pulse - incorporating the
quoted radio error enhances this to $\pm$ 60  $\mu$s. Thus to
first order, the optical trails the X-ray main peak, and the X-ray trails
the $\gamma$-ray main peaks, but their absolute
location with respect to the radio is not clear, being
dependent on the local epoch timing solution. The discrepancies are
undoubtedly as a result of the differing quality of the 
different datasets (in terms of S/N), the accuracy of the radio 
ephemeris solution and  the manner in which the main peak's 
centroid is defined. From our folded light curves,
it appears that the main peak in the $UBV$ is located $after$ the
radio main pulse, but it is important to appreciate that
there is a substantial error on the radio timing solution. What
is required are simultaneous deep observations in both the
optical \& radio so as to guarantee as accurate a timing
solution as possible. Such observations have recently been
made in collaboration
with colleagues from the Westerbork Radio Telescope Observatory 
using the TRIFFID instrument at La Palma; the data
is currently under analysis.
Realistic numerical modelling is possibly the
only full consistent manner in which to reproduce such frequency 
dependent lag effects and so constrain the position(s) of
emission for a standard dipolar structure under a range of viewing
geometries, for both spatially extended \& localised model
frameworks. Such modelling studies are underway at NUI, Galway 
and elsewhere.

\section{Conclusion}

Definitive optical observations of the Crab and other pulsars have to date
been restricted - either by poor temporal resolution both in terms of
resolution and with respect to absolute UTC, and also by the limited
spatial information which is crucial in distinguishing the pulsar's
emission from the background. Using the TRIFFID high speed photometer,
we have obtained datasets uncompromised by these previous limitations,
and in addition to confirming many of the previous conclusions of
Eikenberry et al. (1997), we have determined several new results
uniquely due to the technology we have implemented, including the
fluxing of the unpulsed component of emission, the absolute arrival
times of the three light curves with respect to one another, and
the spatial extent of the plateau on the main peak. Our photometric
analysis was however limited by the reference data we used to
renormalize the flux ratios thus determined. It is clear that
further analysis using more photometrically accurate reference
data combined with further observations at extended wavebands,
such as the $UV$ or $R/I$ would yield fluxes that would provide excellent
leverage to the least-squares fits and thus produce more significant
power-law exponents. In terms of a more rigorous temporal analysis
of the light curves, deeper observations in $UBV$ would place limits
on the extent of the main peak's plateau, and possibly indicate
if it is energy dependent. Furthermore, a greater sample of photons
per pulse period may allow us to probe the possible correlation
between the radio giant pulses and high energy emission mechanisms.
We note that the incorporation of
polarimetric technology into this nascent field of 2-d high
speed photometry would yield critical data which would undoubtedly 
reveal more clues to puzzle over, and thus contribute
towards the eventual solution to this three decade old mystery of
how exactly pulsars function.

\begin{acknowledgements}
{\it 
ESO is thanked for the provision
of their MAMA detector. We are  grateful to the 6-m telescope Program
Committee of the RAS for observing time allocation.  We thank the
engineers of SAO RAS, A.  Maksimov for help in equipment preparation for
the observations and the Director of SAO RAS Yu.  Balega for arranging
the observations. Dr Ray Butler of NUI, Galway is thanked for his 
assistance during the course of the photometric analysis. This work was 
supported by the Russian Foundation of Fundamental Research (98-02-17570), 
State programme "Astronomy", Russian Ministry of
Science and Technical Politics, and the Science-Educational Centre
"Cosmion", and funded under the INTAS programme.
The support of Enterpise Ireland, the Irish Research and Development 
agency, is very gratefully acknowledged. Finally, we would like
to thank the anonymous referee's considerable effort and
wise counsel in contributing towards the final version.
 
}

\end{acknowledgements}

\end{document}